# Liquid-fueled Oblique Detonation Stabilized by a Transverse Jet

Wenhao Wang[a,b,c], Zongmin Hu[a,c,*], and Peng Zhang[b,*]

[a] *State Key Laboratory of High-Temperature Gas Dynamics (LHD), Institute of Mechanics, Chinese Academy of Sciences, Beijing 100190, China*

[b] *Department of Mechanical Engineering, City University of Hong Kong, Kowloon Tong, Kowloon 999077, Hong Kong*

[c] *School of Engineering Sciences, University of Chinese Academy of Sciences, Beijing 100049, China*

---

**Abstract**

The role of a transverse liquid n-heptane jet in initiating and stabilizing liquid n-heptane oblique detonation waves (ODWs) in a confined model combustor was computationally investigated in the present work. The jet-to-inflow momentum ratio, $J$, was identified as the primary control parameter. Under steady inflow pressures, a weak jet with a small $J$ fails to initiate an ODW; a slightly stronger jet ignites only a local near-normal detonation between the OSW and the separation shock wave without forming a developed ODW branch; a moderate jet establishes a standing detonation wave system consisting of an ODW, a near-normal detonation branch, and a separation shock wave; a large but still admissible $J$ produces a wall-coupled ODW-Mach-stem configuration; and an excessive jet momentum destabilizes the ODW by pushing it out of the combustor into the external compression region. Under oscillatory inlet pressure, the standing ODW remains dynamically stabilized within the combustor through bounded, phase-dependent transitions between distinct combustion modes. At sufficiently large $J$, the transverse jet ceases to act as an effective stabilization actuator. The resulting dynamic-stabilization map reveals a finite operating window governed jointly by jet momentum and inlet-pressure fluctuation.

---

*Corresponding author.

Zongmin Hu: huzm@imech.ac.cn

Peng Zhang: penzhang@cityu.edu.hk

## Nomenclature

| **Terminology** | |
|---|---|
| $A_d$ | droplet surface area (m²) |
| $A_p$ | inlet-pressure fluctuation amplitude (-) |
| $B_M$ | Spalding mass-transfer number (-) |
| $c_p$ | liquid heat capacity (J kg⁻¹ K⁻¹) |
| $C_p$ | gas mixture heat capacity (J kg⁻¹ K⁻¹) |
| $C_d$ | drag coefficient (-) |
| $D$ | density-weighted diffusion coefficient (m² s⁻¹) |
| $\bar{\mathcal{D}}_f$ | average binary diffusion coefficient (m² s⁻¹) |
| $D_p$ | droplet diameter (m) |
| $E$ | total specific energy (J kg⁻¹) |
| $e$ | specific internal energy (J kg⁻¹) |
| $\boldsymbol{F}_p$ | droplet force (N) |
| $h_c$ | convective heat-transfer coefficient (W m⁻² K⁻¹) |
| $J$ | jet-to-inflow momentum ratio (-) |
| $m_p$ | droplet mass (kg) |
| $\dot{m}_p$ | droplet evaporation rate (kg s⁻¹) |
| $MW_i$ | molecular weight of species i (kg mol⁻¹) |
| $N$ | number of inlet-pressure cycles (-) |
| $N_p$ | number of parcels in a cell (-) |
| $n_q$ | number of chemical reactions (-) |
| $n_s$ | number of gas species (-) |
| Nu | Nusselt number (-) |
| $p$ | pressure (Pa) |
| $p_0$ | baseline inlet pressure (Pa) |
| $p_{\mathrm{in}}$ | inlet pressure (Pa) |
| $p_s$ | surface/saturation pressure (Pa) |
| Pr | Prandtl number (-) |
| $\dot{Q}_c$ | convective heat-transfer rate (J s⁻¹) |
| $\dot{Q}_l$ | latent heat-transfer rate (J s⁻¹) |
| $Q_+$ | positive heat-release rate |
| $\mathbf{q}$ | diffusive heat-flux vector (W m⁻²) |
| $R$ | specific gas constant (J kg⁻¹ K⁻¹) |
| $R_u$ | universal gas constant (J mol⁻¹ K⁻¹) |
| $Re_p$ | droplet Reynolds number (-) |
| Sc | Schmidt number (-) |
| Sh | Sherwood number (-) |
| $S_m$ | gas-phase mass source term (kg m⁻³ s⁻¹) |
| $S_F$ | gas-phase momentum source term (N m⁻³) |
| $S_e$ | gas-phase energy source term (W m⁻³) |
| $S_{s,i}$ | gas-phase species source term (kg m⁻³ s⁻¹) |
| $S_0$ | wedge-induced OSW reference |
| $S_j$ | jet-induced OSW reference |
| $T$ | gas temperature (K) |
| $T_p$ | droplet temperature (K) |
| $T_s$ | film temperature (K) |
| $t$ | time (s) |
| $\boldsymbol{u}$ | gas velocity vector (m s⁻¹) |
| $u_j$ | liquid-jet injection velocity (m s⁻¹) |
| $\boldsymbol{u}_p$ | droplet velocity vector (m s⁻¹) |
| $u_\infty$ | post-OSW inflow velocity (m s⁻¹) |
| $V_c$ | computational-cell volume (m³) |
| $\mathrm{We}_p$ | droplet Weber number (-) |
| $X_s$ | mole fraction at the droplet surface (-) |
| $Y_i$ | mass fraction of species i (-) |
| $Y_s$ | fuel-vapor mass fraction at the interface (-) |
| $Y_\infty$ | far-field fuel-vapor mass fraction (-) |

| ***Greek letters*** | |
|---|---|
| $\Delta t$ | pressure-forcing window (s) |
| $\Delta\tau_e$ | signed convective delay relative to Sj (s) |
| $\Delta x_{t_n}$ | triple-point displacement (m) |
| $\mu$ | dynamic viscosity (Pa s) |
| $\nu$ | kinematic viscosity (m² s⁻¹) |
| $\rho$ | gas density (kg m⁻³) |
| $\rho_j$ | liquid-jet density (kg m⁻³) |
| $\rho_p$ | liquid/droplet density (kg m⁻³) |
| $\rho_s$ | gas density at droplet surface (kg m⁻³) |
| $\rho_\infty$ | post-OSW inflow density (kg m⁻³) |
| $\boldsymbol{\tau}$ | deviatoric stress tensor (Pa) |
| $\tau_{\mathrm{hr}}$ | heat-release timescale proxy (s) |
| $\tau_{\mathrm{res}}$ | remaining convective residence time (s) |
| $\dot{\omega}_i$ | net production rate of species i (kg m⁻³ s⁻¹) |

| **Abbreviations** | |
|---|---|
| CFL | Courant-Friedrichs-Lewy number |
| DW | detonation wave |
| KNP | Kurganov-Noelle-Petrova scheme |
| LPT | Lagrangian particle tracking |
| LTO | low-temperature oxidation |
| MS | Mach stem |
| NDW | near-normal detonation wave |
| ODW | oblique detonation wave |
| ODWE | oblique detonation wave engine |
| OSW | oblique shock wave |
| PDF | probability density function |
| PSI-CELL | Particle-Source-In-Cell |
| R-R | Rosin-Rammler |
| RSW | reflected shock wave |
| SL | streamline |
| SSW | separation shock wave |

## 1. Introduction

Oblique detonation waves (ODWs), consisting of a shock-supported reaction front attached to a compression surface, have long been regarded as a compelling combustion mode for hypersonic air-breathing propulsion. This is because they can, in principle, realize rapid heat release with high thermodynamic efficiency in a compact flow path [1-5]. Developing practical liquid-fueled oblique detonation wave engines (ODWEs), however, demands far more than merely identifying a theoretical ODW solution. The fundamental challenge lies in initiating and anchoring a standing detonation inside a finite combustor subjected to realistic aerodynamic disturbances [6, 7]. This difficulty becomes sharper when geometric confinement interacts with a liquid-hydrocarbon spray-air inflow. Under such gas-liquid conditions, wave reflection, interphase momentum and heat transfer, flow separation, and localized shock interactions jointly control detonation onset, confinement, and persistence.

The wedge-induced ODW configuration has long provided a canonical setting for examining the fundamental physics of ODW initiation and stabilization. Early studies focused mainly on the formation criteria for steady ODWs, using detonation-polar analysis and idealized post-shock heat-release models. Later numerical and experimental studies of gaseous mixtures showed that initiation is intrinsically unsteady, often involving transverse-wave interactions, thermochemical instabilities, and complex structural evolution [8-11]. Recent work has moved beyond idealized, unconfined wedges and exposed the stringent constraints imposed by combustor confinement. Zhang et al. [12] demonstrated experimentally that the combustor geometry must be adjusted to selectively establish theoretical strong or weak detonation modes. In confined configurations, multi-wave interactions and wave-wall coupling control the global wave structure. Wang et al. [13] showed that the merger of subsonic regions generated by the Mach stem and downstream shock reflections governs the steadiness of the ODW complex. Viscous effects add another destabilizing pathway: the adverse pressure gradient produced by wave-wall interactions can induce severe boundary-layer separation, and the resulting detonation Mach stem may cause geometric flow choking, force the wave system upstream, and destabilize the entire combustor-scale structure [14-16].

To overcome these technical barriers, considerable research has been dedicated to active and passive flow-control strategies. These approaches fall mainly into three categories: geometric modification, local energy deposition, and fluidic actuation. Geometric methods, including double-wedge interactions [17], stepped surfaces [18], multi-stage wedges [19], and specific on-wedge trips [7, 20], enhance localized compression and promote earlier initiation. Dynamic geometric control, such as proportional wedge-angle adjustment, has also been proposed to stabilize the autoignition point [21, 22]. Laser-induced hot spots provide a different route by accelerating local ignition through direct energy deposition [23]. Fluidic actuation offers a more flexible and readily implementable alternative because the injected mass flow and momentum can be adjusted without changing the combustor geometry. Hot jets [24-26], cold jets [27], sweeping jets [28], and combined jet-wedge configurations [29] have shown that localized aerodynamic forcing can reshape the compression field, shorten the ignition distance, and shift the position at which the detonation front stabilizes.

Most control strategies were developed for gaseous mixtures, particularly hydrogen/air mixtures, where actuation mainly targets gas-phase shock-chemistry coupling. Liquid hydrocarbons introduce additional difficulties to the problem: a multiscale two-phase ignition process in which droplet breakup, evaporation, vapor redistribution, and finite-rate chemistry are strongly coupled [7]. Ren et al. [30] showed that two-phase ODW stabilization differs from gaseous stabilization because droplets may evaporate incompletely before reaching the reaction zone. They further demonstrated that the initiation length is governed by the competition between evaporative cooling and chemical heat release, leading to a distinctive $\Lambda$-shaped dependence on equivalence ratio [31]. These effects can produce wave systems, such as double-ODW structures, that differ markedly from hydrogen-air systems [32]. Recent studies of $n$-heptane sprays linked autoignition, mode transition, and wave detachment to droplet loading, evaporation history, reaction-zone structure, and the synchronization between evaporation and gas-phase heat release [33-36]. For n-dodecane spray-air mixtures, Wang et al. [37] showed that increasing flight altitude weakens breakup, enlarges the shock-reaction-front separation, and shifts initiation from an abrupt to a smooth mode. These studies clarify the multiphase physics of liquid-fueled ODWs, yet active control of their initiation and stabilization remains underexplored. This gap is central to practical ODWEs, which must both initiate a two-phase detonation and hold it within a confined combustor under evolving fuel-delivery and inflow conditions.

A transverse liquid jet offers a physically compatible control strategy for liquid-fueled ODWEs. Aerodynamically, it reshapes the local shock system and shortens the ODW initiation distance; thermochemically, it enhances droplet breakup, accelerates evaporation, and enriches the mixture near the jet-induced wave interaction. Our recent study showed that transverse liquid jets can initiate liquid-fueled ODWs over an unconfined wedge, with the controlling mechanism shifting from chemically assisted ignition to physically driven wave forcing as jet momentum increases [38]. In a confined combustor with unsteady inflow, the jet must not only trigger an ODW but also sustain a standing wave within a complex reflected-wave system while the inlet pressure oscillates. Although pre-evaporated multiphase ODWs can adjust their fronts dynamically to accommodate upstream pressure oscillations [39, 40], the response of a transverse-jet-controlled liquid-fueled ODW under simultaneous confinement and oscillatory inflow remains unresolved.

The present study computationally investigates transverse-jet-controlled liquid-fueled ODWs in a confined combustor under steady and oscillatory inflow conditions, using a validated Eulerian-Lagrangian two-phase detonation solver. For steady inflow, we determined how the jet-to-inflow momentum ratio governs ignition, standing-wave stabilization, droplet evaporation, and combustor-scale mode transitions. For oscillatory inlet pressure, we tracked the dynamic response of the jet-controlled wave system and identified the ranges of jet momentum and forcing amplitude over which detonative combustion remains confined. These analyses clarify the dual role of transverse liquid injection as both an ignition promoter and a dynamic stabilization actuator for liquid-fueled ODWEs.

The rest of the paper is organized as follows. Section 2 describes the numerical methodology, physical models, and computational setup. Section 3 examines jet-induced ignition and combustion-mode transitions under steady inflow, followed by the dynamic response and stability window under oscillatory inlet pressure. Section 4 summarizes the main conclusions.

## 2. Formulation and computational methodology

### *2.1. Governing equations and sub-models*

We adopted an Eulerian-Lagrangian framework for the compressible, chemically reacting flow of n-heptane droplets dispersed in a gaseous carrier. In the Eulerian frame, the continuous phase obeys the multi-component compressible Navier-Stokes equations,

$$\frac{\partial \rho}{\partial t} + \nabla \cdot (\rho \boldsymbol{u}) = S_m, \tag{1}$$

$$\frac{\partial (\rho \boldsymbol{u})}{\partial t} + \nabla \cdot [\boldsymbol{u}(\rho \boldsymbol{u})] + \nabla p - \nabla \cdot \boldsymbol{\tau} = \boldsymbol{S}_F, \tag{2}$$

$$\frac{\partial (\rho E)}{\partial t} + \nabla \cdot [\boldsymbol{u}(\rho E)] + \nabla \cdot (\boldsymbol{u} p) + \nabla \cdot \boldsymbol{q} - \nabla \cdot (\boldsymbol{\tau} \cdot \boldsymbol{u}) = S_e, \tag{3}$$

$$\frac{\partial (\rho Y_i)}{\partial t} + \nabla \cdot [\boldsymbol{u}(\rho Y_i)] + \nabla \cdot [-D \nabla (\rho Y_i)] = \dot{\omega}_i + S_{s,i}, (i = 1,2, \dots, ns - 1). \tag{4}$$

In Eqns. (1)-(4), $\rho$, $\boldsymbol{u}$, and $p$ are the gas density, velocity vector, and pressure, respectively. The equation of state is $p = \rho R T$, and the mixture gas constant is $R =$ $R_u \sum_{i=1}^{n_s} Y_i / \mathrm{MW}_i$. Here, $R_u = 8.314\ \mathrm{J/(mol \cdot K)}$ is the universal gas constant, $MW_i$ is the molecular weight of species $i$, $n_s$ is the number of species, and $Y_i$ is its mass fraction. Finite-rate chemistry contributes the net production rate $\omega_i$, which is evaluated from a 44-species, 112-reaction skeletal mechanism for n-heptane [41], validated extensively against experimental data for n-heptane-air detonations [34, 42, 43]. The Newtonian stress tensor is $\boldsymbol{\tau} = \mu[\nabla \boldsymbol{u} + (\nabla \boldsymbol{u})^T - 2(\nabla \cdot \boldsymbol{u})\mathbf{I}/3]$, with $\mu$ evaluated from Sutherland's law. The total specific energy and heat flux are $E = e + |\boldsymbol{u}|^2/2$ and $\boldsymbol{q} = -k\nabla T$, where $k$ is the thermal conductivity. Under the unity-Lewis-number approximation, the density-weighted diffusion coefficient is $D = k/(\rho C_p)$, with the mixture heat capacity $C_p = \sum_{i=1}^{n_s} Y_i C_{p,i}$. The source terms $S_m$, $\boldsymbol{S}_F$, $S_e$, and $S_{s,i}$ account for interphase exchange of mass, momentum, energy, and species.

The dispersed phase is advanced in a Lagrangian framework. The trajectory, mass, and thermal state of each droplet follow the Lagrangian particle tracking (LPT) equations,

$$\frac{dm_d}{dt} = \dot{m}_p, \tag{5}$$

$$\frac{d\boldsymbol{u}_p}{dt} = \frac{\boldsymbol{F}_p}{m_p}, \tag{6}$$

$$c_p \frac{dT_p}{dt} = \frac{\dot{Q}_c + \dot{Q}_l}{m_p}. \tag{7}$$

Here, $m_p$, $\boldsymbol{u}_p$, and $T_p$ denote the droplet mass, velocity, and temperature. With the droplets assumed spherical, $m_p = \pi \rho_p D_p^3/6$; $\rho_p$ is the liquid density and $D_p$ is the droplet diameter. The liquid specific heat at constant pressure is $c_p$. The terms $\dot{m}_p$, $\boldsymbol{F}_p$, $\dot{Q}_c$, and $\dot{Q}_l$ represent, respectively, evaporative mass loss, aerodynamic drag, convective heat transfer, and latent heat transfer.

Droplet evaporation is evaluated with the Abramzon-Sirignano model [44]. The droplet mass-depletion rate is $\dot{m}_p = -\pi D_p \, \mathrm{Sh} \, \bar{\mathcal{D}}_f \rho_s \ln(1 + B_M)$, where $\bar{\mathcal{D}}_f$ is the average binary diffusion coefficient in the gas film. The gas density at the interface follows $\rho_s = p_s MW_m/(R_u T_s)$; the film temperature is $T_s = (T + 2T_p)/3$, and $p_s$ is obtained from a polynomial function of $T_s$ [45]. Convective mass transport is described by $\mathrm{Sh} = 2.0 + 0.6\,\mathrm{Re}_p^{1/2}\mathrm{Sc}^{1/3}$, with $\mathrm{Re}_p = \rho D_p \mid \boldsymbol{u}_p - \boldsymbol{u} \mid/\mu$ and $\mathrm{Sc} = \mu/(\rho \bar{\mathcal{D}}_f)$ defining the droplet Reynolds and Schmidt numbers. The Spalding mass-transfer number is $B_M = (Y_s - Y_\infty)/(1 - Y_s)$, where $Y_\infty$ is the far-field fuel-vapor mass fraction. The interfacial fuel-vapor mass fraction is $Y_s = MW_p X_s / \sum_i (X_i MW_i)$, while Raoult's law gives $X_s = X_l p_{\mathrm{sat}}(T_p)/p$; the saturation pressure $p_{\mathrm{sat}}$ is evaluated as a polynomial function of $T_p$ [45].

Momentum exchange is closed using the spherical-droplet drag relation $\boldsymbol{F}_p = -(18\mu/\rho_p D_p^2)(C_d \mathrm{Re}_p/24)\, m_p(\boldsymbol{u}_p - \boldsymbol{u})$, where $C_d$ is the drag coefficient. The droplets experience high-Reynolds-number conditions in the present detonation flow; for $\mathrm{Re}_p > 1000$, we take $C_d = 0.424$ [46]. Zhu et al. [47] showed that this constant-drag approximation differs from more detailed correlations, such as the Brown-Lawler model, by no more than 6.4%, which is adequate for the present supersonic momentum-exchange calculation.

Heat exchange contains sensible convection and evaporation-induced latent cooling. The convective contribution is $\dot{Q}_c = h_c A_d (T - T_p)$, where $A_d$ is the droplet surface area, $h_c = \mathrm{Nu}\, k/D_p$ is the heat-transfer coefficient, and $\mathrm{Nu} = 2.0 +$ $0.6\,\mathrm{Re}_p^{1/2}\mathrm{Pr}^{1/3}$ is the Ranz-Marshall Nusselt number [48], with $\mathrm{Pr} = c_p\mu/k$. The latent contribution is written as $\dot{Q}_l = \dot{m}_p h_l(T_p)$ ; $h_l(T_p)$ denotes the latent heat of vaporization at the droplet temperature. Because evaporation gives $\dot{m}_p < 0$, this term acts as a cooling contribution in the droplet energy equation. The Ranz-Marshall correlation is a standard convective-transport closure in two-phase detonation solvers, validated extensively by Huang et al. [49] and adopted widely in recent studies [30, 47, 50].

Droplet breakup is represented by the Pilch-Erdman model [51], which has been employed widely in liquid-fueled detonation simulations [33, 34, 36, 43, 52]. The model separates breakup into five experimentally based regimes and assigns a characteristic breakup time to each. Its Weber-number coverage extends from $O(10)$ to values above $10^3$ , spanning the droplet deformation and fragmentation states encountered in the present spray-detonation calculations.

The gas and dispersed phases are coupled through the Particle-Source-In-Cell (PSI-CELL) method [53]. Parcel-level exchange rates are accumulated onto the Eulerian control volume containing each parcel. For a cell of volume $V_c$ containing $N_p$ parcels, the gas-phase source terms are

$$S_m = -\frac{1}{V_c}\sum_{1}^{N_p}\dot{m}_p\,, \tag{8}$$

$$\boldsymbol{S}_F = -\frac{1}{V_c}\sum_{1}^{N_p}\boldsymbol{F}_p\,, \tag{9}$$

$$S_e = -\frac{1}{V_c}\sum_{1}^{N_p}\left(\dot{Q}_c + \dot{Q}_l\right), \tag{10}$$

$$S_{s,i} = -\frac{1}{V_c}\sum_{1}^{N_p}\dot{m}_p\,\delta_{iF}. \tag{11}$$

Here, $\delta_{iF} = 1$ for the condensed species and $\delta_{iF} = 0$ for all other species. The leading minus signs convert droplet-centered exchange rates into the corresponding gas-phase source terms with the appropriate sign.

The coupled equations are integrated with the authors' in-house, open-source two-phase detonation solver developed within the *rhocentralfoam* framework of OpenFOAM V7 [54]. The solver combines finite-rate chemistry, multicomponent transport, and Lagrangian droplet dynamics and has been validated against theoretical solutions and experimental data for gaseous and liquid-fueled ODW problems; its two-phase transverse-jet capability has also been benchmarked against experimental measurements in our previous work [20, 38, 54]. Time integration uses a first-order Euler scheme, and spatial derivatives are evaluated with the Gauss-limited Linear scheme. The gas-phase CFL number is kept below 0.3. Shock waves, contact discontinuities, and sharp detonation fronts are captured with the Kurganov-Noelle-Petrova central-upwind scheme [55].

### *2.2. Problem configuration and numerical specifications*

Figure 1 shows the model ODWE configuration and the two-dimensional computational domain used in the simulations. Because the present study targets the initiation and stabilization of liquid-fueled ODWs by a transverse liquid jet, together with the controlling mechanisms, a two-dimensional domain is sufficient for this purpose. Three-dimensional jet-inflow interaction, including the associated vortical and wave-system structures, does influence the flow field; given the present scope and the computational cost, however, such effects are left to future work. The engine adopts an external-compression layout for a Mach-9 inflow, with two successive 12.5° forebody compression stages that guide the incoming stream toward a confined combustor. The computational domain, shown in Fig. 1(b), resolves the combustor region downstream of the inlet plane. A 25° compression ramp generates the ODW, and a transverse injector is placed 50 mm downstream of the combustor lip. The jet radius is $R_j$ = 5 mm, and the jet issues normally from the wall into the incoming *n*-heptane spray-air mixture. The transverse spray follows a Rosin-Rammler size distribution with a mean droplet diameter of 10 μm.

The inflow prescribed at the computational inlet is a premixed two-phase *n*-heptane/air stream. The transverse injector supplies additional liquid fuel and locally modifies the shock structure, droplet-processing region, and heat-release distribution. At the combustor entrance, the inflow temperature and velocity are 814.4 K and 2466 m/s, respectively. For the steady cases, the inlet pressure is fixed at $p_0$= 1.963 × $10^5$ Pa. These simulations examine how the jet-to-inflow momentum ratio governs the transition from ignition failure to a standing detonation and, ultimately, to loss of stabilization within the combustor. For the unsteady cases, the inlet pressure is sinusoidally modulated about the baseline value as

$$p_{\text{in}}(t) = p_0\left[1 + A_p \sin(2\pi N t/\Delta t)\right], \tag{12}$$

where $A_p$ is the nondimensional pressure-fluctuation amplitude, $\Delta t$ = 1 ms, and $N$ = 4 or 8. Thus, the imposed inlet pressure completes either four or eight oscillation cycles within the 1 ms computational window. Each sinusoidal case represents a single Fourier mode of an arbitrary temporal inlet disturbance, so the mode-by-mode response characterized here supplies the elementary building block from which the reaction to broadband forcing can be reconstructed. The present sinusoidal analysis of the jet behavior and the initiation-stabilization mechanism provides the basis on which more complex, spatially varying inflow distortion can be investigated in future work.

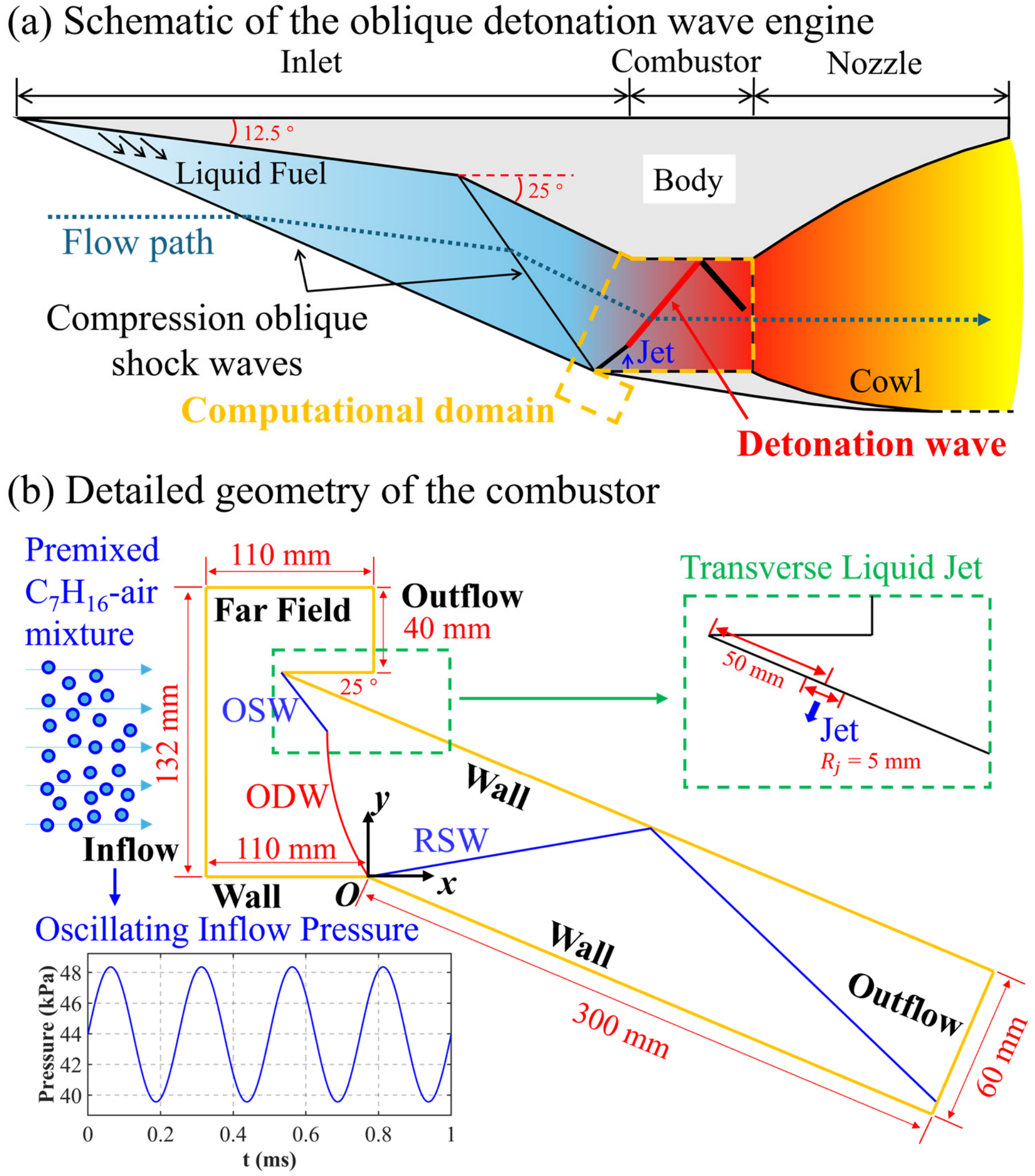

Fig. 1. Schematic of the transverse-liquid-jet-controlled oblique detonation wave engine (ODWE) and the computational setup. (a) External-compression engine layout for Mach-9 inflow, with two successive 12.5° forebody compression stages directing the spray–air stream into a confined combustor; the dashed yellow box marks the computational domain. (b) Detailed geometry of the combustor and computational domain, showing the 25° compression ramp, the transverse n-heptane liquid jet ($R_j$ = 5 mm) located 50 mm downstream of the combustor lip, the inflow/outflow boundaries, and the principal wave structures. The inset plots the sinusoidal inlet-pressure forcing imposed in the unsteady simulations. OSW: Oblique shock wave; ODW: Oblique detonation wave; RSW: Reflection shock wave.

*2.3. Grid-independence Study*

A grid-independence study was performed for the two-phase ODW simulations using three uniform grid spacings: 0.2, 0.3, and 0.4 mm. All cases employ the 44-species, 112-reaction skeletal n-heptane mechanism introduced in Section 2.1. Figure 2 compares the pressure, temperature, and gas-phase $n$-heptane mass fraction, $Y_{C_7H_{16}}$, along the same representative streamline. The three grids predict nearly identical shock locations, post-shock temperature levels, pressure evolution, and fuel-vapor profiles. Minor differences appear only near sharp compression and reaction fronts, where the finest grid resolves slightly steeper gradients. The 0.3 mm grid captures the principal wave-induced jumps and thermochemical evolution while reducing the computational cost relative to the 0.2 mm grid. It was therefore used for all subsequent simulations.

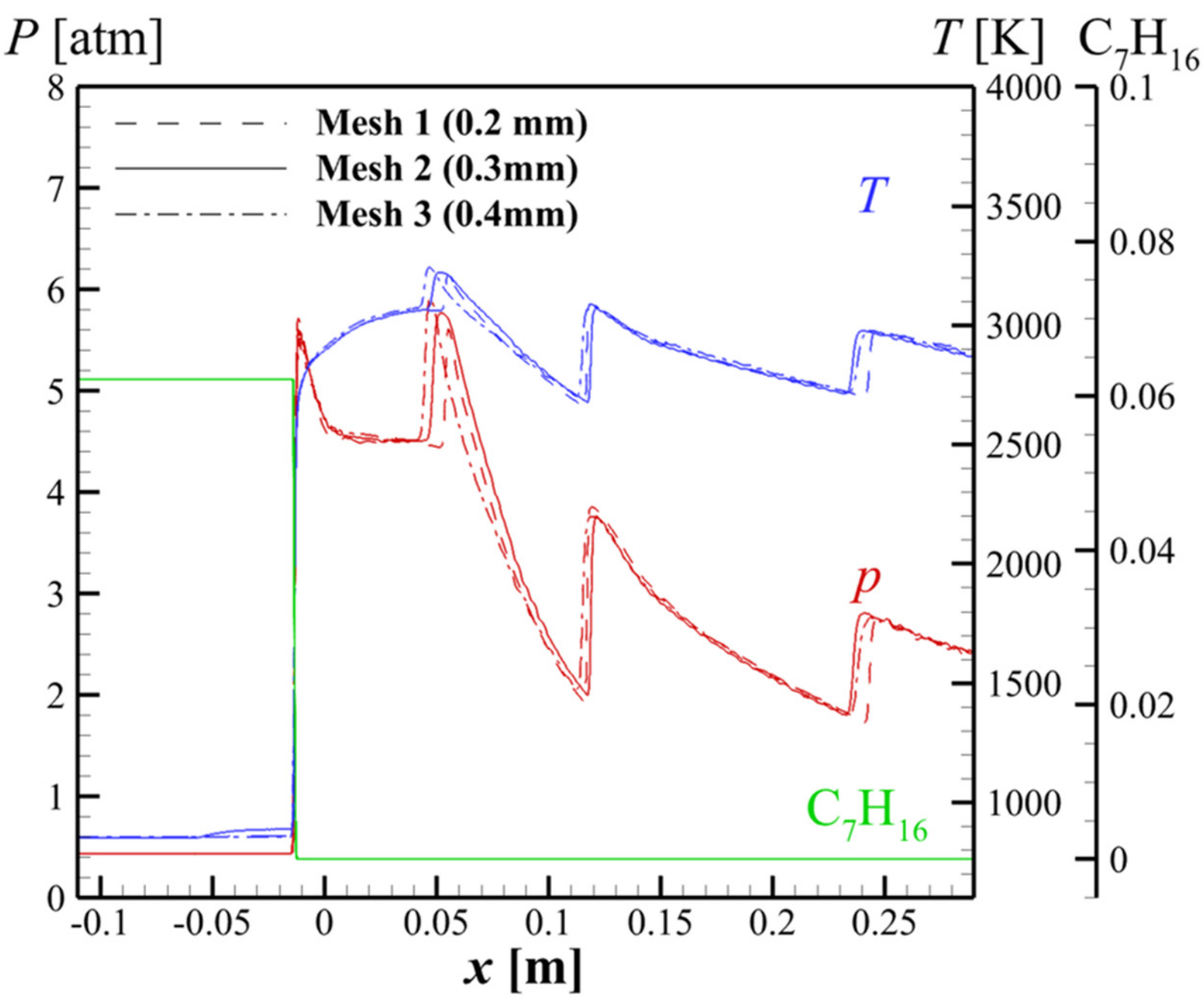

Fig. 2. Grid-independence study along a representative streamline for $\Delta x = 0.2$, 0.3, and 0.4 mm, showing pressure $p$, temperature $T$, and gas-phase $n$-heptane mass fraction $Y_{C_7H_{16}}$.

## 3. Results and Discussion

### 3.1. *Effect of transverse liquid jets under steady inflow pressure*

#### 3.1.1. *Wave structure and combustion modes*

The jet intensity is quantified by the momentum ratio, $J = \rho_j u_j^2 / \rho_\infty u_\infty^2$, where $\rho_j$ and $u_j$ are the liquid-jet density and injection velocity, and $\rho_\infty$ and $u_\infty$ denote the density and velocity of the post-oblique-shock inflow. Specifically, $J$ is adjusted by varying the injection velocity $u_j$ of the transverse liquid $n$-heptane jet. Without transverse injection, the baseline $n$-heptane spray-air inflow does not initiate detonation [20]. A finite transverse liquid jet removes this initiation barrier and establishes standing detonative combustion inside the combustor, providing computational evidence that transverse liquid injection can both initiate and sustain a gas-liquid two-phase ODW in a confined configuration.

Figure 3 presents nine flow fields spanning the $J$-sweep at $t$ = 1 ms, ordered by increasing jet momentum from $J$ = 0.01 to $J$ = 2.0. Each panel shows the gas-temperature field overlaid with the instantaneous droplet distribution, wave structures, and, where appropriate, an enlarged view of the principal wave-interaction region. Together, the nine cases trace a continuous progression through five distinct combustor-scale regimes: non-ignition ($J$ = 0.01), a locally standing OSW-NDW-SSW structure without a developed ODW branch ($J$ = 0.02, 0.03), a standing OSW-ODW-NDW-SSW combustor mode ($J$ = 0.04 - 0.5), a wall-coupled OSW-ODW-MS configuration ($J$ = 1.0), and loss of stabilization within the combustor under excessive jet momentum ($J$ =

2.0). The intermediate cases ($J$ = 0.03, 0.04, 0.05, 0.5) demonstrate that the mode transitions proceed continuously rather than through isolated selected states.

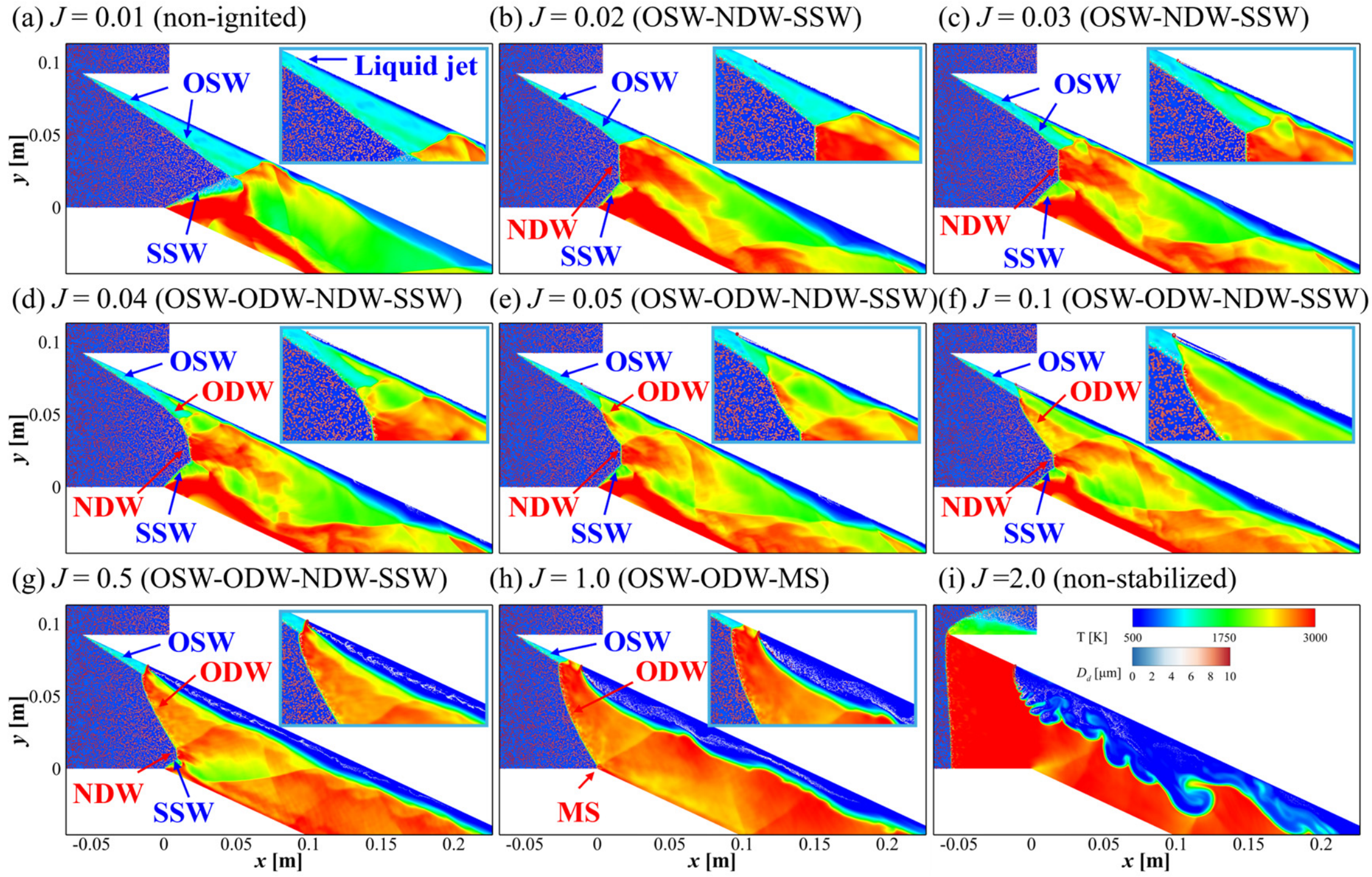

Fig. 3. Steady wave-mode transition driven by the jet-to-inflow momentum ratio $J$ at $t$ = 1 ms. Each panel shows the instantaneous gas-temperature field overlaid with the droplet distribution and wave-structure annotations. The sequence proceeds from (a) non-ignition at $J = 0.01$, to (b, c) a locally standing OSW–SSW field with a developing near-normal detonation wave but no established ODW branch at $J = 0.02$ and 0.03, to (d–g) a standing OSW-ODW–NDW–SSW combustor mode over $0.04 \leq J \leq 0.5$, to (h) a wall-coupled OSW-ODW–MS configuration at $J = 1.0$, and finally to (i) loss of stabilization within the combustor at $J = 2.0$. Insets in (a) – (h) enlarge the jet–shock interaction region. Common color bars for the gas temperature $T$ and droplet diameter $D_d$ are shown in (i).

At $J = 0.01$, the jet-induced oblique shock wave (OSW) remains too weak to raise the compressed spray-air mixture above the autoignition threshold. The flow is governed mainly by the OSW and the separation shock wave (SSW), but no attached high-temperature reaction zone forms immediately behind the shock system. This state therefore corresponds to a non-ignited regime.

As $J$ increases to 0.02 and 0.03, the jet-induced OSW still cannot directly establish a sufficiently strong ODW. The larger interaction angle between the OSW and the SSW, however, creates a near-normal detonation wave (NDW) in the lower interaction region. Although this NDW remains locally standing, the combustor has not yet reached the desired oblique-detonation mode because a fully developed ODW branch is still absent. Comparing panels (b) and (c) shows that the NDW region grows and sharpens as $J$ is increased within this sub-regime, but the upper portion of the wave system remains a non-reactive OSW.

Ignition of the ODW first emerges at $J$ = 0.04. With further increase to $J$=0.1, the ODW couples with the SSW and forms an intermediate NDW in the lower interaction region, producing a standing ODW-NDW-SSW combustor mode. The strengthened jet-induced compression steepens the reaction front and displaces the high-temperature zone upstream, showing that the transverse liquid jet no longer merely preconditions the droplets but actively reorganizes the wave structure. As demonstrated by panels (d) - (g) of Fig. 3, this combustor mode persists over a broad range of approximately 0.04 $\leq J \leq$ 0.5: the wave topology is preserved, while the ODW angle, the NDW height in the lower interaction region, and the size of the high-temperature zone all evolve continuously with $J$.

At higher jet momentum, $J = 1.0$, the ODW steepens further, and the near-wall reflection develops into a stable wall-coupled Mach stem (MS). The SSW no longer acts as the dominant lower-wall stabilization structure. Instead, the ODW-MS configuration produces a larger and more compact high-temperature region, consistent with a stronger post-detonation state.

Beyond the standing operating window, the $J = 2.0$ case shows that excessive jet momentum drives the detonation front upstream of the combustor stabilization region. The flow field becomes dominated by an extended high-temperature zone and a strongly distorted droplet-depleted layer. The wave system can therefore no longer remain stabilized by the combustor-scale OSW/SSW interaction, marking loss of stabilization within the combustor.

### *3.1.2. Wave-mode-dependent droplet distribution*

Figure 4 connects the first four wave structures in Fig. 3 with the corresponding droplet statistics. The left column defines three sampling regions on the instantaneous flow field. Region I samples the incoming droplets before strong wave processing. Region II follows droplets that have primarily passed through the wedge-induced OSW and its associated compression/shear layer. Region III samples the droplets convected into the downstream complex wave system, where the local structure changes from the OSW/SSW-dominated field to the ODW-NDW-SSW and ODW-MS configurations. The middle column plots the droplet diameter-temperature distribution, with gray points denoting all droplets and colored symbols marking the three regions. The right column gives the corresponding diameter PDFs.

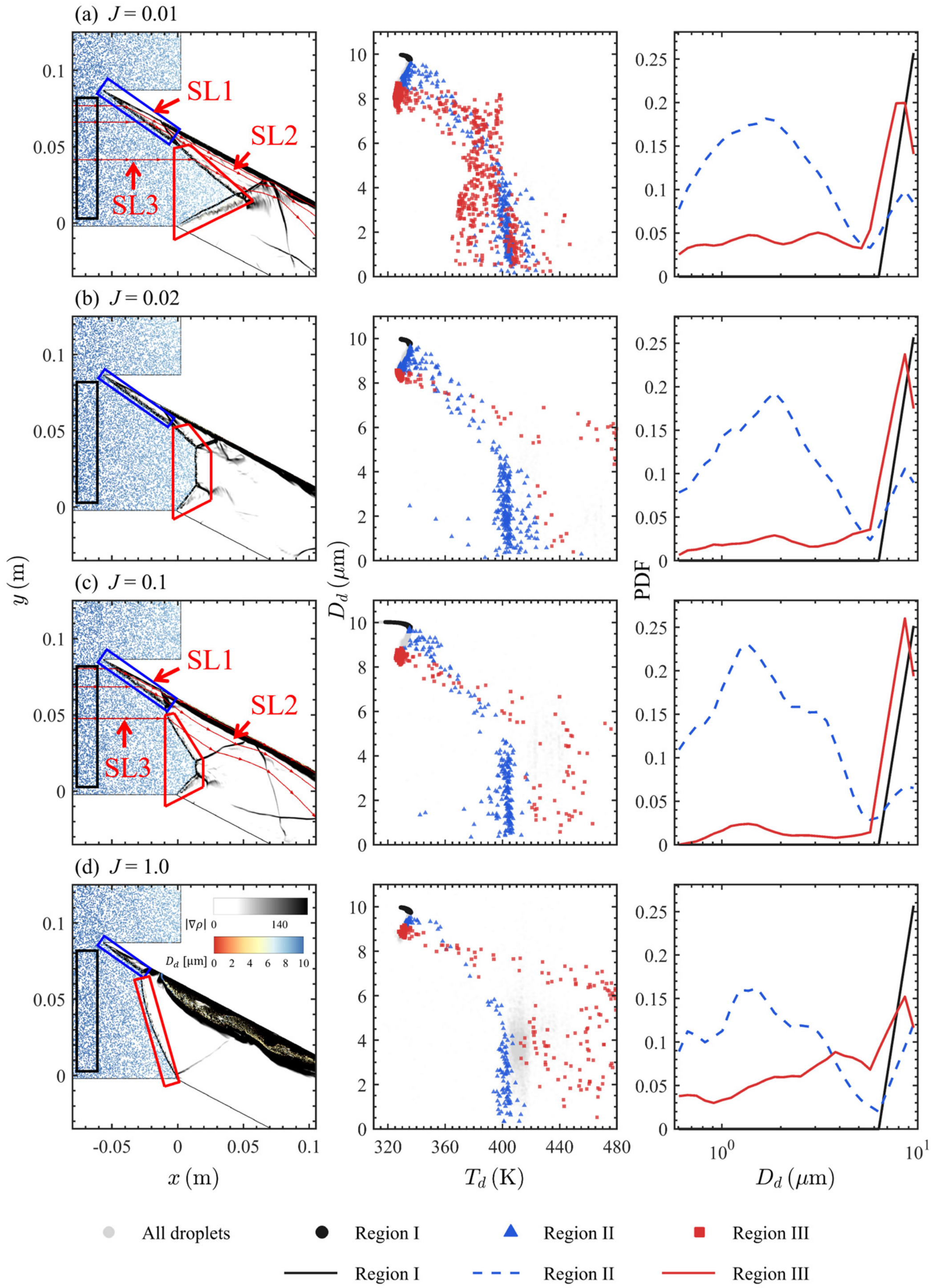

Fig. 4. Wave-mode-dependent droplet statistics for four representative steady cases: (a) $J = 0.01$, (b) $J = 0.02$, (c) $J = 0.1$, and (d) $J = 1.0$. The left column defines three sampling regions on the instantaneous density-gradient field. Region I samples incoming droplets upstream of the wave system, Region II follows droplets that have passed mainly through the wedge-induced OSW and the jet-driven compression/shear layer, and Region III samples droplets convected into the downstream complex wave system. The middle column plots droplet diameter $D_d$ against droplet temperature $T_d$, with gray points denoting all droplets and colored symbols marking the three regions. The right column gives the corresponding diameter probability density functions.

Across all four cases, Region I remains clustered near the initial droplet diameter and the lowest droplet temperature, and its PDF stays concentrated near the injected size. These compact black points provide a baseline, confirming that the changes in Regions II and III arise from wave-induced processing rather than inlet variability. Region II consistently shifts toward smaller diameters, showing that passage through the OSW produces the first major heating and size-reduction event. Region III exhibits the strongest case-to-case variation. Its distribution records how each downstream wave mode processes the droplets. Thus, increasing $J$ does not simply make all droplets smaller; it reorganizes where, and under which wave structure, the droplets are heated and evaporated.

At $J = 0.01$, Region II already contains a broad small-diameter population, showing that the primary OSW can heat and downsize part of the incoming droplet cloud. Region III, however, still contains many heated droplets and retains a pronounced large-diameter branch in the PDF. Thus, although weak shock processing and partial evaporation occur, the downstream wave system does not immediately generate a coupled detonative reaction zone that can rapidly evaporate, consume, and reorganize the droplet population. This behavior is consistent with the OSW/SSW-dominated non-ignited mode in Fig. 3(a).

At $J = 0.02$, the separation between the OSW-processed droplets and the downstream reaction-zone droplets becomes clearer. Region II clusters at reduced diameters after OSW passage, whereas Region III extends toward higher droplet temperatures and includes droplets sampled along the newly formed NDW/SSW interaction. The spatial and statistical separation between these two populations matches the local normal-detonation structure shown in Fig. 3(b).

At $J = 0.1$, the ODW-NDW-SSW structure produces a stronger separation between the post-OSW droplets and the downstream reaction-zone droplets. Region II remains dominated by small droplets generated by OSW-driven heating and evaporation, and its diameter peak shifts to smaller values than in the $J = 0.02$ case. This shift reflects the strengthening of the primary OSW by the transverse jet, which intensifies droplet vaporization before the flow reaches the main reaction region. Region III also contains more low-diameter, high-temperature droplets than in Fig. 4(b), mainly because the newly established ODW exposes the dispersed phase to stronger and more spatially organized detonation heating. At the same time, fewer droplets reach the highest-temperature range above approximately 460 K, which is consistent with the reduced dominance of the localized NDW as the ODW branch takes over the main heat-release structure.

At $J = 1.0$, the stronger transverse jet substantially steepens the ODW, and the detonation structure occupies much of the combustor. The Region II diameter peak decreases further and the population becomes sparser, indicating that the stronger jet shortens the effective OSW-processing length and brings the initiation region upstream. Region III spans a broad high-temperature range and contains droplets with both intermediate and reduced diameters, while its PDF gains weight at small and intermediate sizes compared with the lower-$J$ cases. Increasing $J$, therefore, does not simply reduce every droplet diameter monotonically. It changes the wave structure and initiation location, thereby altering the downstream residence history, strengthening droplet heating and vaporization, and reinforcing the ODW-supported reaction structure.

### *3.1.3. Chemical kinetics and timescale analysis*

The droplet statistics show where the dispersed phase is heated, downsized, and evaporated, but they do not by themselves determine whether detonation is initiated. Ignition also requires fuel-vapor production, low-temperature oxidation, and main heat release to couple within the finite interaction length of the jet-induced shock system. Figure 5 examines this coupling for two representative cases: the non-ignited case at $J = 0.01$ and the standing-ODW case at $J = 0.1$. For each case, three streamlines are extracted — SL1 passes closest to the jet-interaction region, SL2 samples a path farther from the jet influence, and SL3 lies near the central combustor region. Figure 5(a, b) follows the streamline-resolved reaction history along SL1 for the non-ignited and ignited cases, respectively. The upper panels show the normalized gas-phase n-heptane, $O_2$, and OH profiles, while the lower panels show the combined low-temperature-oxidation (LTO) pool defined as

$$Y_{\mathrm{LTO}} = Y_{\mathrm{CH_2O}} + Y_{\mathrm{HO_2}} + Y_{\mathrm{H_2O_2}} + Y_{\mathrm{PC_7H_{15}O_2}} + Y_{\mathrm{OC_7OOH}}, \quad (13)$$

together with the normalized positive heat-release rate $\dot{Q}_+$ and the temperature $T$.

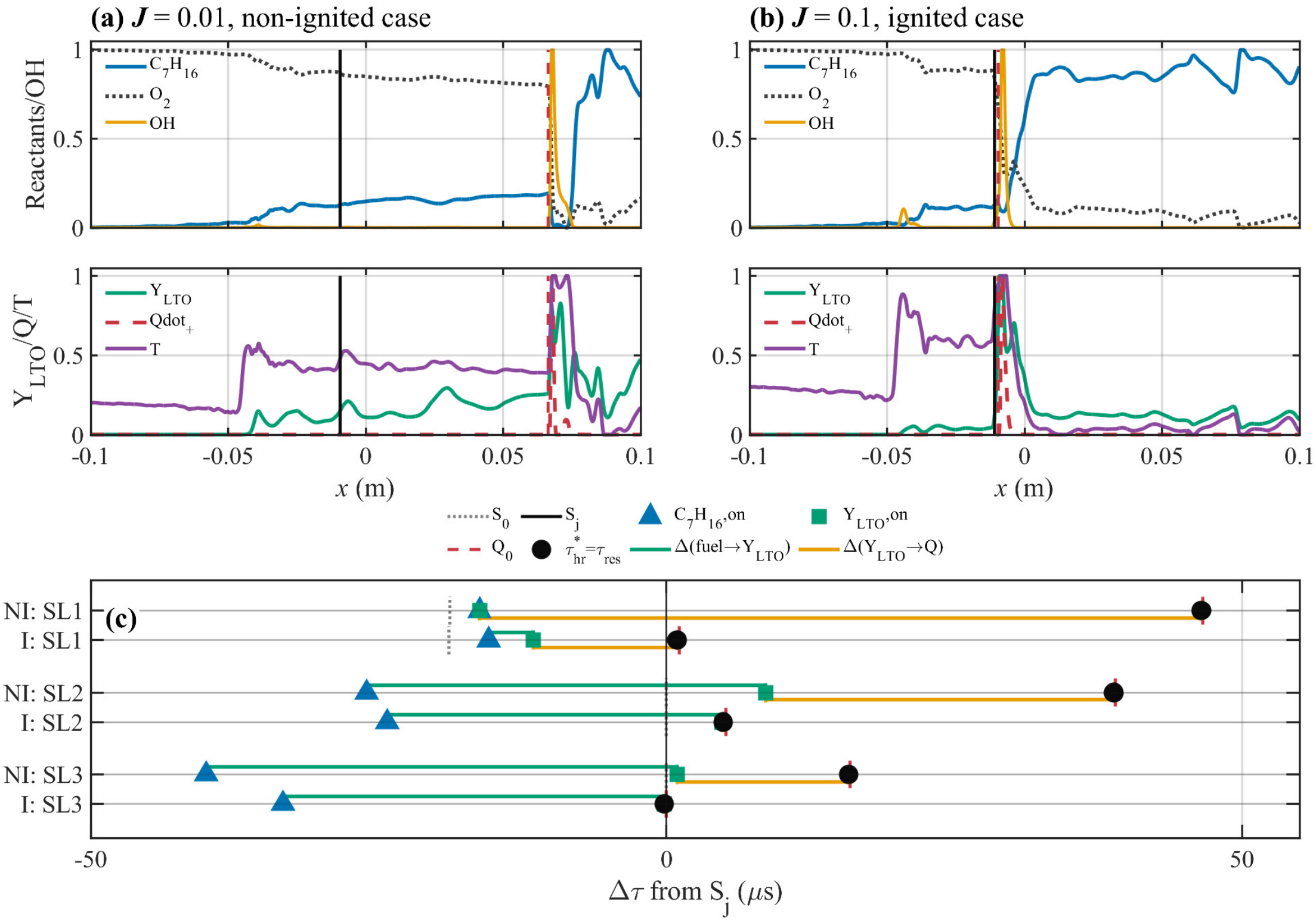

Fig. 5. Streamline-resolved reaction sequencing and shock-referenced timing alignment for the non-ignited case, $J = 0.01$ (NI), and the ignited standing-ODW case, $J = 0.1$ (I). (a, b) Reaction histories along SL1, the streamline passing closest to the jet-induced wave-interaction region, for (a) the non-ignited case and (b) the ignited case. The upper panels show the normalized gas-phase n-heptane, $O_2$, and OH profiles; the lower panels show the low-temperature-oxidation pool, $Y_{LTO} = Y_{CH_2O} + Y_{HO_2} + Y_{H_2O_2} + Y_{PC_7H_{15}O_2} + Y_{OC_7OOH}$, together with the normalized positive heat-release rate, $Q_+$, and temperature, $T$. Vertical markers denote the wedge-induced OSW reference, $S_0$, the jet-induced OSW reference, $S_j$, and the heat-release onset, $Q_0$. (c) Shock-referenced timing of fuel-vapor production, low-temperature oxidation, and main heat release for three streamlines, plotted as signed convective delays, $\Delta\tau$, relative to $S_j$. Triangles mark the fuel-vapor onset, $C_7H_{16,on}$; squares mark the LTO onset, $Y_{LTO,on}$; and black circles mark the location where the heat-release timescale proxy, $\tau^*_{hr}$, equals the remaining convective residence time, $\tau_{res}$. The green and yellow bars show the intervals from fuel-vapor onset to LTO onset and from LTO onset to main heat release, respectively.

At $J = 0.01$, the OSW has already produced the small-diameter Region II population, but the chemical response along SL1 remains delayed. The jet-induced shock raises the temperature and gradually promotes fuel-vapor accumulation and LTO-species formation; however, the main heat-release event remains far downstream of the jet-induced OSW reference position, $S_j$. The peaks of OH and $\dot{Q}_+$ appear only after a long induction distance, by which point the wave system can no longer sustain a standing ODW inside the combustor. The non-ignited case is therefore not caused by the absence of droplet processing or chemistry. Rather, the weak jet cannot couple compression, evaporation, LTO chemistry, and main heat release within the finite interaction length available in the combustor.

At $J = 0.1$, the reaction sequence changes markedly. The Region II and Region III populations in Fig. 4(c) show that droplets have passed through both the OSW-processing region and the downstream ODW-NDW-SSW structure. Along SL1, fuel-vapor growth and the LTO pool rise before, or immediately around, $S_j$, followed by a sharp increase in $\dot{Q}_+$ and $\mathrm{OH}$ as $\mathrm{O_2}$ is rapidly consumed. This compressed ordering indicates that low-temperature oxidation has prepared a reactive mixture before the flow enters the stronger compression region. The main heat release therefore couples directly with the ODW-NDW interaction instead of being convected far downstream, which explains the successful establishment of a standing oblique-detonation mode.

To compare the reaction sequence among different streamlines, Fig. 5(c) maps each event location onto a signed convective delay relative to the local jet-induced OSW, $S_j$. For an event $e$ on a given streamline, the signed delay is defined as

$$\Delta\tau_e = \tau_e - \tau_{S_j}, \tag{14}$$

where the convective time is obtained by integrating $ds/u$ along the streamline and $u$ is the velocity magnitude. Thus, $\Delta\tau_e < 0$ indicates that the event occurs upstream of $S_j$, whereas $\Delta\tau_e > 0$ indicates that it occurs downstream of the jet-induced OSW. The earlier shock marker, $S_0$, is retained as a reference for the wedge-induced OSW. The gas-phase fuel onset, $C_7H_{16,\mathrm{on}}$, and the LTO onset, $Y_{\mathrm{LTO,on}}$, are detected from the smoothed $C_7H_{16}$ and $Y_{\mathrm{LTO}}$ profiles using a fixed fraction of their streamline-local maxima. The heat-release onset, $Q_0$, is defined from the smoothed positive heat-release signal, $\dot{Q}_+ = \max(\dot{Q}, 0)$. The black circle marks the first location at which the heat-release timescale proxy becomes comparable to the remaining convective residence time. The heat-release timescale proxy is defined as

$$\tau_{\mathrm{hr}}^* = \frac{p}{\dot{Q}_+}, \tag{15}$$

where $p$ is the local pressure and $\dot{Q}_+$ is the positive volumetric heat-release rate. Since $\dot{Q}_+$ has the dimensions of pressure per unit time, $\tau_{\mathrm{hr}}^*$ estimates the time required for heat release to modify the local pressure-scale energy density. The remaining convective residence time is defined as

$$\tau_{\mathrm{res}} = \int_{s_0}^{s_{\mathrm{e}}} \frac{ds}{u}, \tag{16}$$

where $s_{\mathrm{e}}$ denotes the downstream end of the analysis window. The crossing $\tau_{\mathrm{hr}}^* = \tau_{\mathrm{res}}$ marks the point at which local heat release becomes fast enough to influence the flow within the remaining convective time available along the streamline.

Figure 5(c) applies this shock-referenced timing analysis to three streamline samples in the non-ignited $J = 0.01$ case and the ignited $J = 0.1$ case. In the non-ignited case, $Q_0$ and the $\tau_{\mathrm{hr}}^* = \tau_{\mathrm{res}}$ marker remain displaced downstream of $S_j$ for all three streamlines. Strong heat release therefore develops only after the flow has passed the effective jet-induced wave-interaction region. In the ignited case, by contrast, these heat-release markers collapse toward $S_j$, and the interval from $Y_{\mathrm{LTO,on}}$ to $Q_0$ becomes much shorter. The fuel-to-$Y_{\mathrm{LTO}}$ interval remains more streamline-dependent, especially for SL2 and SL3, consistent with the mixed Region III droplet population in Fig. 4(c). The decisive difference is therefore not fuel-vapor production alone, but whether the final transition from low-temperature preparatory chemistry to high-temperature heat release occurs within, or very close to, the jet-induced compression region.

These results show that jet-induced initiation requires vapor production, low-temperature chemistry, and main heat release to be spatially compact and closely aligned within the jet-induced compression region. The steady-inflow cases therefore establish the baseline stabilization mechanism: the transverse liquid jet must be strong enough to produce sufficient droplet heating, breakup, and vaporization, and to synchronize $Y_{\mathrm{LTO}}$ accumulation and main heat release with the downstream detonative interaction. Excessive jet momentum, however, over-steepens the wave system and destroys the global stabilization. The oscillatory-inflow cases examined below test whether this coupled wave-droplet-chemistry organization remains bounded when the inlet pressure varies in time.

### *3.2. Effect of transverse liquid jets under oscillating inflow pressure*

#### *3.2.1. Phase-dependent mode switching and heat-release locking during one pressure cycle*

Figure 6 compares the instantaneous wave structures over one imposed inlet-pressure cycle for three jet-momentum ratios, $J = 0.1$, 0.5, and 1.0, at the common forcing condition $A_p = 0.5$ and $N = 4$. The six rows correspond to $t = 0.75$, 0.80, 0.85, 0.90, 0.95, and 1.00 ms, which span one full pressure-oscillation period. From left to right, the columns show how the jet-stabilized ODW responds as the jet momentum increases from the interior of the standing window ($J = 0.1$), to its margin ($J = 0.5$), and then to an over-forced state ($J = 1.0$).

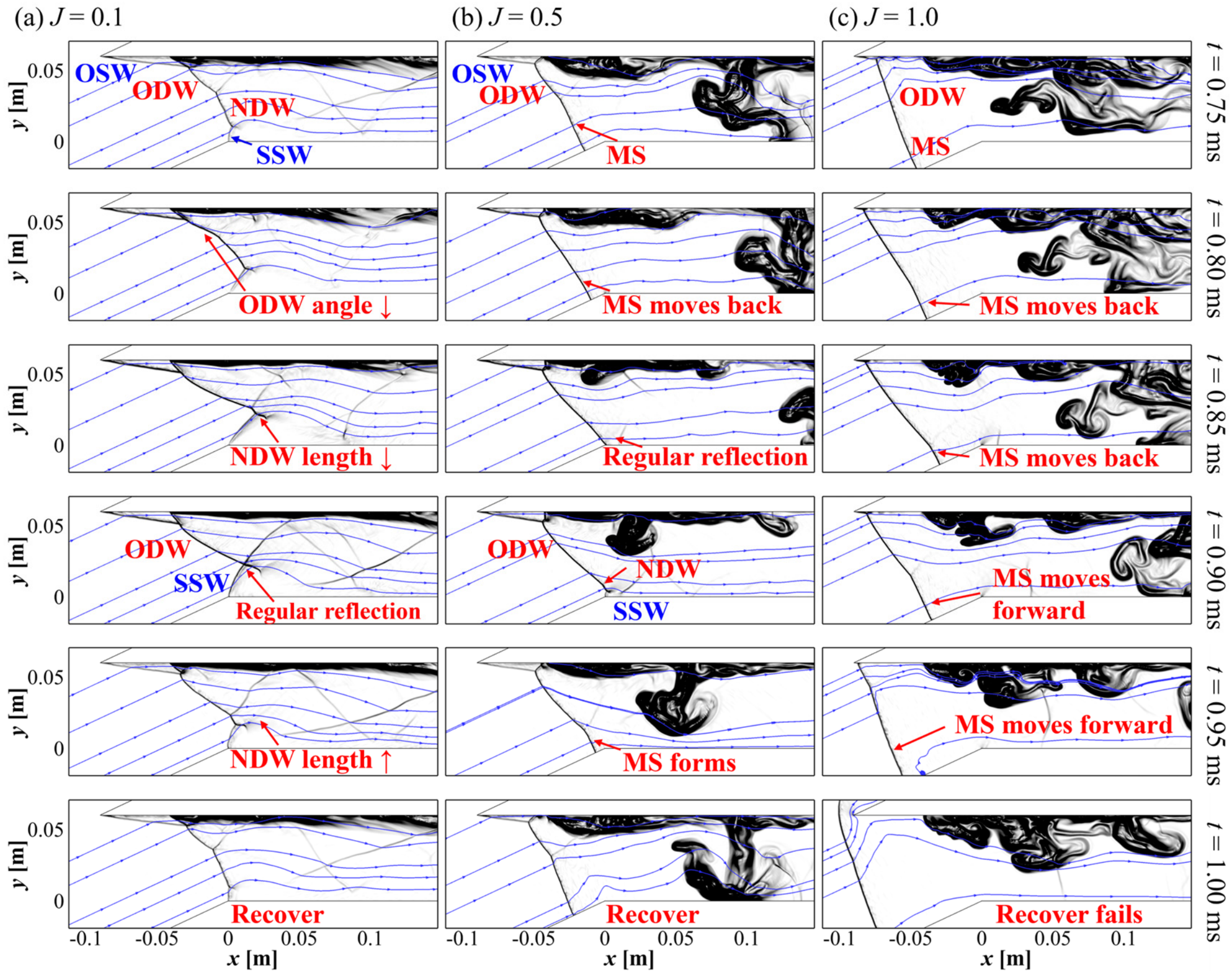

Fig. 6. Phase-dependent wave-mode switching over one full inlet-pressure oscillation cycle for three jet-momentum ratios: (a) $J = 0.1$, (b) $J = 0.5$, and (c) $J = 1.0$. All cases use $A_p = 0.5$and $N = 4$. Each column shows numerical-schlieren snapshots overlaid with streamlines at six instants, $t =$ 0.75, 0.80, 0.85, 0.90, 0.95, and 1.00 ms. At $J = 0.1$, the combustor remains dynamically stabilized through bounded transitions between ODW–NDW–SSW and ODW–SSW structures and recovers its starting topology by the end of the cycle. At $J = 0.5$, the cycle still closes, but recovery proceeds through a broader sequence involving a wall-coupled ODW–MS state, regular reflection, and an intermediate ODW–NDW–SSW topology. At $J = 1.0$, the Mach stem drifts forward during the pressure-recovery half-cycle, and the initial configuration is not restored, indicating loss of dynamic stabilization.

At $J = 0.1$, Fig. 6(a) shows a dynamically stabilized cycle. At $t = 0.75$ ms, the combustor contains an OSW-ODW-NDW-SSW structure: the jet-induced OSW compresses the incoming two-phase mixture, and the ODW couples to the lower-wall SSW through a Mach-reflection-like interaction, producing an intermediate NDW in the lower interaction region. As the inlet pressure falls, the ODW angle decreases at $t = 0.80$ ms, and the NDW shortens at $t = 0.85$ ms. The reflection then degenerates into regular reflection, leaving an ODW-SSW structure at $t = 0.90$ ms. During the pressure-recovery half-cycle, the NDW grows again at $t = 0.95$ ms, and the original OSW-ODW-NDW-SSW structure is recovered by $t = 1.00$ ms. Thus, although the imposed pressure oscillation modulates both the ODW angle and the near-wall reflection, the wave system remains phase-locked to the forcing and returns to its initial state within one cycle.

At $J = 0.5$, Fig. 6(b) lies closer to the edge of the standing window and displays a stronger structure response. The cycle begins at $t = 0.75$ ms with an OSW-ODW-MS structure, in which a wall-coupled Mach stem replaces the smaller NDW branch observed at $J = 0.1$. As the inlet pressure decreases, the Mach stem retreats downstream at $t = 0.80$ ms, and the reflection collapses into regular reflection at $t = 0.85$ ms. By $t = 0.90$ ms, the wave system has reorganized into an ODW-NDW-SSW structure. When the inlet pressure rises again, the Mach stem re-forms at $t = 0.95$ ms, and the initial OSW-ODW-MS structure is restored by $t = 1.00$ ms. The combustor therefore remains dynamically stabilized, but recovery now proceeds through a broader sequence of intermediate states, including regular reflection and an ODW-NDW-SSW configuration.

At $J = 1.0$, Fig. 6(c) shows that the same forcing no longer produces a closed cycle. The wave system initially forms an OSW-ODW-MS structure at $t = 0.75$ ms. During the pressure-decrease half-cycle, the Mach stem retreats downstream at $t = 0.80$ and 0.85 ms, resembling the early response at $J = 0.5$. Once the inlet pressure begins to recover, however, the Mach stem does not return to its original position. Instead, it drifts forward, toward the combustor inlet, at $t = 0.90$ and 0.95 ms. By $t = 1.00$ ms, the wave system has shifted substantially upstream, and the initial OSW-ODW-MS structure is not recovered. The final panel therefore marks a loss of dynamic stabilization rather than a delayed phase response.

Figure 6 therefore shows that oscillatory inflow does not merely perturb a fixed detonation structure. It drives phase-dependent transitions among OSW-ODW-SSW, OSW-ODW-NDW-SSW, and OSW-ODW-MS states. At moderate jet momentum, these transitions remain bounded and reversible; at excessive jet momentum, the same inlet-pressure modulation produces a non-recoverable forward drift of the Mach stem and the attached detonation structure. The transverse liquid jet therefore controls not only the mean wave structure but also the recoverability of the liquid-fueled ODW under unsteady inflow.

The visual sequence in Fig. 6 suggests that recoverability depends not only on the instantaneous wave topology but also on whether heat release remains locked to the jet-induced compression. To test this connection quantitatively, we plotted in Figure 7(a) the convective-time offset, $\tau_{S_j} - \tau_{Q_0}$, between the jet-induced OSW reference, $S_j$, and the heat-release onset, $Q_0$, for the same three cases considered in Fig. 6. Positive values indicate that $Q_0$ occurs earlier than $S_j$, corresponding to an upstream-detached heat-release onset. At $J = 0.1$ and 0.5, this offset remains close to zero throughout the forcing cycle. The main heat-release event therefore stays tied to the jet-induced compression region, even as the inlet pressure modulates the ODW angle and the near-wall reflection. This phase locking explains why the wave systems in columns (a) and (b) of Fig. 6 recover their initial configurations after one cycle.

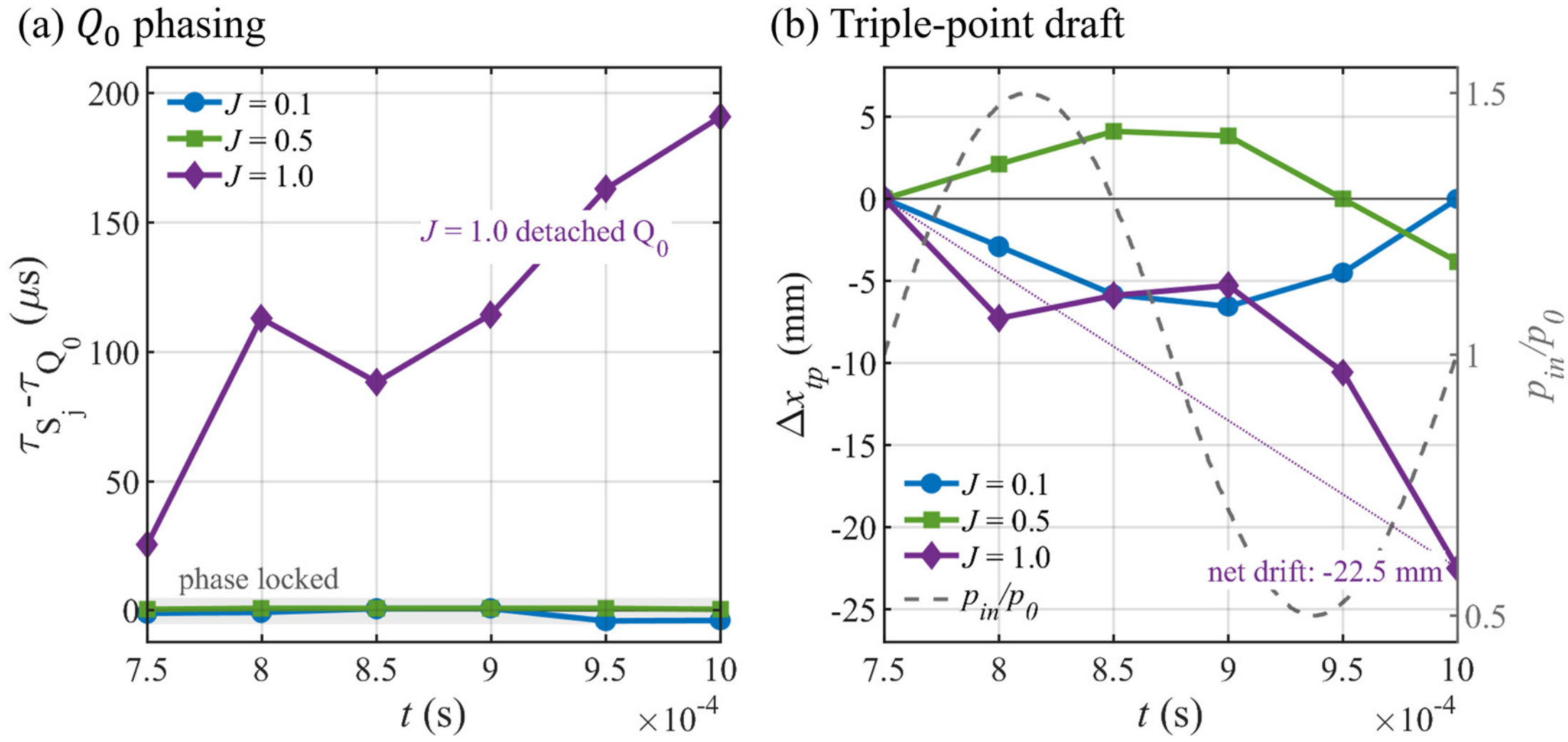

Fig. 7. Temporal phasing and triple-point drift under oscillatory inlet-pressure. Panels (a) and (b) compare the phase-resolved response under the same forcing condition, $A_p = 0.5$ and $N = 4$, but with different jet momentum ratios, $J = 0.1$, 0.5, and 1.0. Panel (a) shows the temporal offset between the jet-induced shock $S_j$ and the heat-release onset $Q_0$, quantified by $\tau_{S_j} - \tau_{Q_0}$. Near-zero values indicate phase-locked ignition, whereas large positive values indicate an upstream-detached $Q_0$. Panel (b) shows the corresponding triple-point displacement, $\Delta x_{\mathrm{tp}} = x_{\mathrm{tp}}(t) - x_{\mathrm{tp}}(t_0)$. The gray dashed curve denotes the normalized inlet pressure, $p_{\mathrm{in}}/p_0$. The low-$J$ cases remain largely phase locked with bounded triple-point motion, while $J = 1.0$ exhibits persistent $Q_0$ detachment and drift-dominated triple-point displacement.

At $J = 1.0$, the behavior changes sharply. The heat-release onset detaches upstream of $S_j$, and the offset grows from approximately 25 $\mu$s at $t = 0.75$ ms to nearly 200 $\mu$s at $t = 1.00$ ms. The high-$J$ case therefore loses dynamic stabilization because the strengthened wave system disrupts the spatial and temporal alignment between jet-induced compression and main heat release. The compact coupling among vapor production, low-temperature chemistry, and heat release, established under steady ignition in Section 3.1.3, is no longer maintained during the unsteady cycle.

Figure 7(b) shows the kinematic consequence of this loss of locking. The phase-locked cases, $J = 0.1$ and 0.5, undergo bounded triple-point motion, with $\Delta x_{\mathrm{tp}}(t) = x_{\mathrm{tp}}(t) - x_{\mathrm{tp}}(t_0)$ remaining within a few millimeters and returning close to its initial value by the end of the cycle. The $J = 0.5$ case retains a small residual displacement, consistent with its marginal character in Fig. 6. In contrast, the phase-detached $J = 1.0$ case develops a drift-dominated response: the triple point moves steadily upstream and reaches a net displacement of about $-22.5$ mm. Thus, once the heat-release onset no longer remains locked to the jet-induced OSW, the combustor cannot correct the triple-point excursion within one forcing period.

### *3.2.2. Amplitude-frequency response and dynamic stabilization window*

Figure 8 examines how the bounded response changes with forcing amplitude and frequency once the system remains within, or close to, the dynamically stabilized regime. The phase portraits are plotted in the $(p_{\mathrm{in}}/p_0, \Delta x_{\mathrm{tp}})$ plane at fixed $J = 0.5$, which lies near the edge of the standing window identified above. These trajectories condense the time-dependent triple-point motion into a recoverability map: a closed loop indicates that the wave system returns to its initial state after one pressure cycle, whereas an open or drifting trajectory would signal loss of dynamic stabilization.

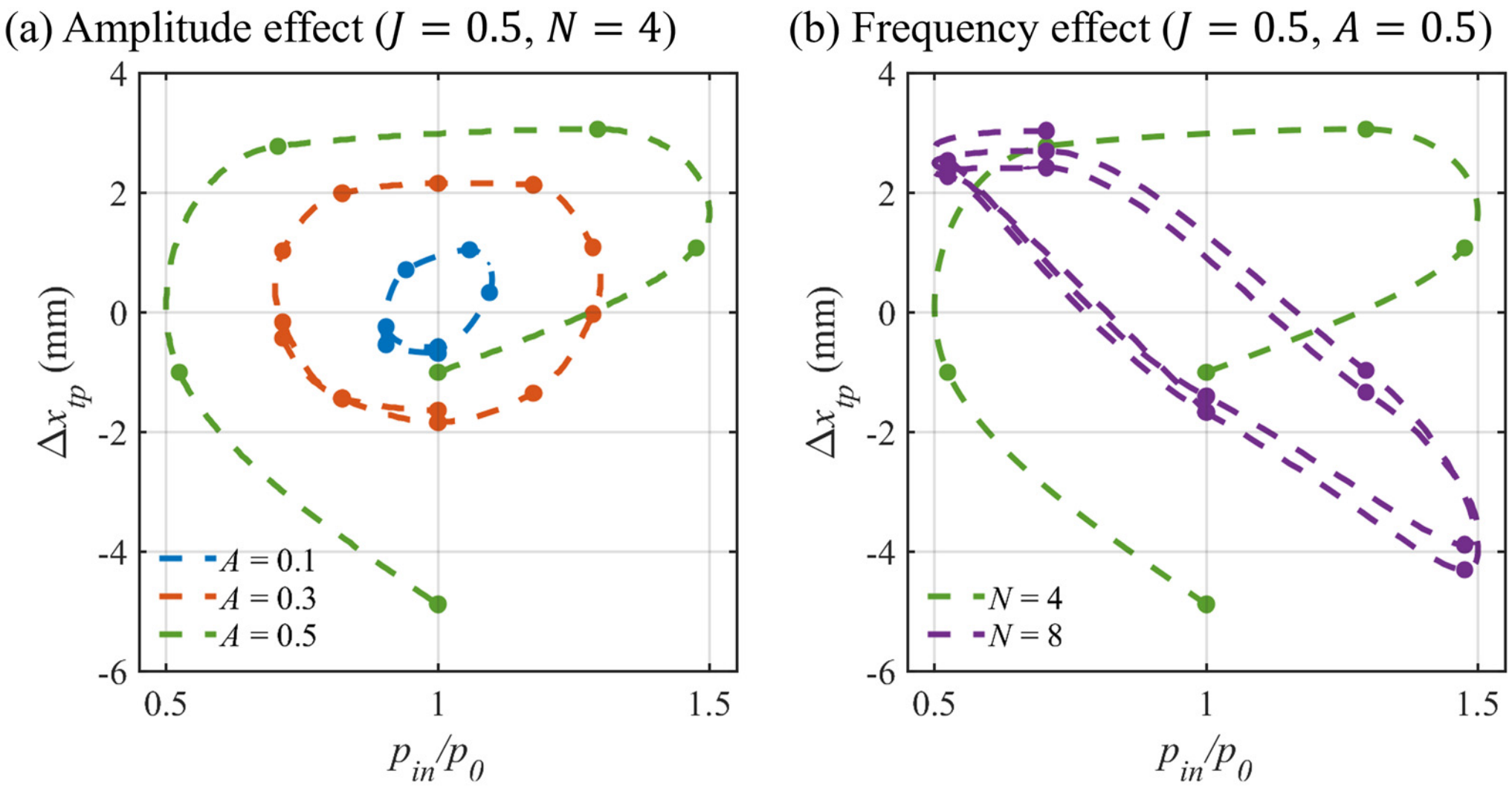

Fig. 8. Amplitude and frequency dependence of the triple-point response under oscillatory inlet-pressure, shown as phase portraits in the $(p_{\mathrm{in}}/p_0, \Delta x_{\mathrm{tp}})$ plane. (a) Amplitude effect at fixed $J = 0.5$ and $N = 4$: increasing $A_p$ from 0.1 to 0.5 enlarges the closed trajectory and increases the triple-point excursion, but the response remains bounded over the sampled amplitudes. (b) Frequency effect at fixed $J = 0.5$ and $A_p = 0.5$: increasing the cycle number from $N = 4$ to $N = 8$ reshapes and tilts the phase trajectory rather than simply amplifying or damping the motion. Markers denote extracted triple-point positions, and dashed curves guide the temporal evolution over one pressure-forcing cycle.

At fixed $N = 4$, increasing the forcing amplitude from $A_p = 0.1$ to 0.5 enlarges the closed phase trajectory, as shown in Fig. 8(a). For $A_p = 0.1$, the triple point remains stabilized to a sub-millimeter-scale excursion around its mean position. At $A_p = 0.3$, the loop expands but remains compact and closed. At $A_p = 0.5$, the excursion reaches approximately $+3$ to $-5$ mm, and the loop spans nearly the full imposed pressure range. The nested structure of the three loops shows that the pressure amplitude can control the magnitude of the triple-point excursion.

At fixed $A_p = 0.5$, changing the forcing cycle number from $N = 4$ to $N = 8$ alters the shape of the phase portrait rather than simply rescaling it, as shown in Fig. 8(b). The $N = 8$ trajectory becomes narrower and more strongly tilted in the $(p_{\mathrm{in}}/p_0, \Delta x_{\mathrm{tp}})$ plane. For the same instantaneous inlet pressure, the triple-point position now depends more strongly on whether the pressure is rising or falling. This behavior reflects the finite response time of the coupled shock-droplet-reaction system: the wave cannot adjust quasi-statically to rapid pressure modulation, and the phase lag becomes part of the stabilizing dynamics. Nevertheless, the $N = 8$ trajectory remains closed. Within the present sampling, higher-frequency forcing modifies the recovery path but does not immediately eliminate dynamic stabilization.

Figure 9 summarizes the sampled dynamic-stabilization behavior in the $(J, A_p)$ plane for $N = 4$. The map should be read as an indicative regime diagram, not as a sharply resolved neutral boundary. At low and moderate jet momentum, the ODW remains dynamically stabilized over the tested pressure amplitudes. The stabilized cases at $J = 0.1$, 0.25, and 0.5 show that transverse liquid injection can sustain bounded wave motion under substantial inlet-pressure forcing, provided that the heat-release onset remains coupled to the jet-induced compression. The case $J = 0.5$, $A_p = 0.5$, is marginal: the wave system remains stabilized, but the enlarged phase loop in Fig. 8(a) and the small residual displacement in Fig. 7(b) place it close to the regime boundary.

At $J = 1.0$, all tested amplitudes lead to loss of stabilization. This result shows that increasing $J$ does not monotonically improve the combustor response. Moderate jet momentum stabilizes the wave system by generating sufficient compression, accelerating droplet processing, and keeping heat release locked to the wave-interaction region. Excessive jet momentum, however, over-steepens the wave system, drives the triple point upstream, and breaks the recoverable coupling between compression and heat release. The dynamic-stabilization window is therefore finite: the transverse liquid jet stabilizes the liquid-fueled ODW only when its momentum is large enough to initiate and anchor the wave, but not so large that it forces the coupled detonation structure upstream and out of the stabilized operating regime.

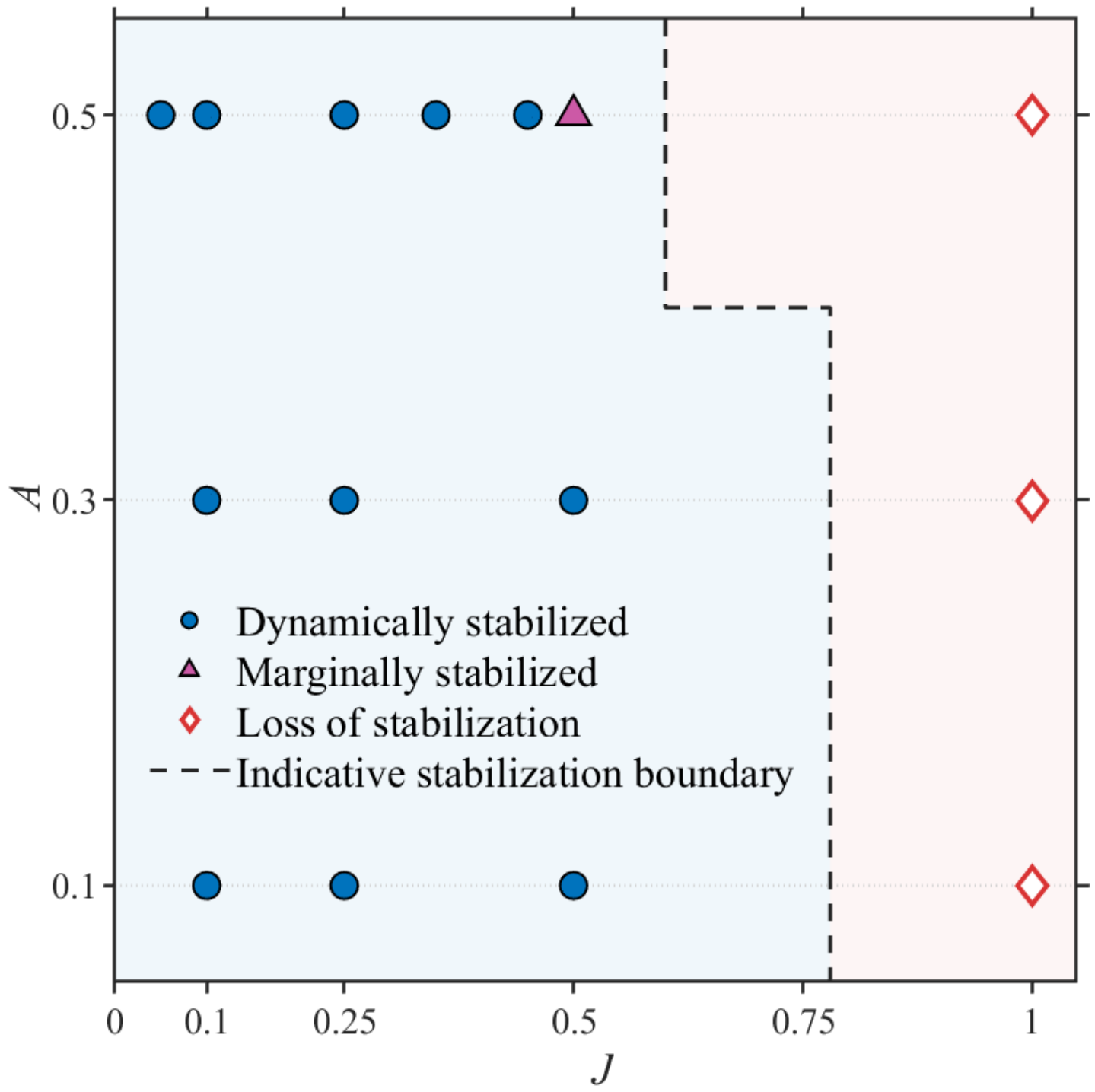

Fig. 9. Sampled dynamic-stabilization map in the jet-momentum-ratio–pressure-fluctuation-amplitude parameter space, $(J, A_p)$, at $N = 4$. Filled blue circles denote cases in which the ODW system remains confined within the combustor, the filled magenta triangle denotes the marginally confined case, and open red diamonds denote unconfined or unstable cases. The dashed line marks an indicative regime boundary; the blue and red shaded regions denote the dynamically confined and unconfined operating envelopes, respectively. Moderate jet momentum sustains bounded mode switching over the tested pressure amplitudes, whereas excessive jet momentum, represented here by $J \approx 1.0$, destroys dynamic confinement for all sampled $A_p$.

## 4. Conclusions

This study computationally investigated the initiation and stabilization of liquid n-heptane ODWs controlled by a transverse liquid n-heptane jet. The results identify the dual role of the liquid jet as both a detonation initiator and a stabilization actuator. Under steady inflow conditions, the jet-to-inflow momentum ratio, $J$, drives the combustor through five distinct wave regimes. Increasing $J$ sweeps the combustor along a single ordered path through three resolved wave structures: a locally standing OSW-NDW-SSW field, in which only a near-normal detonation branch ignites between the jet-induced shock and the separation shock; a fully developed standing OSW-ODW-NDW-SSW configuration coupled to the lower-wall interaction; and a wall-coupled OSW-ODW-MS mode in which the near-wall reflection is folded into a Mach-stem structure. This sequence is bounded by an ignition threshold below which no detonation forms and an expulsion threshold above which the wave system is driven out of the combustor.

The initiation and stabilization mechanism of the transverse liquid jet couples wave structure, droplet/shock interaction, and chemical timescale alignment. The transverse jet reorganizes the combustor-scale wave structure and shifts the relative positions of droplet vaporization, low-temperature chemistry, and the transition to main heat release. Streamline diagnostics for the non-ignited case at $J = 0.01$ and the standing wave case at $J = 0.1$ isolate the controlling step: in the standing case, fuel-vapor onset, LTO chemistry, and main heat release collapse toward the jet-induced OSW interaction, whereas in the non-ignited case these stages remain dispersed downstream of the effective wave-interaction region.

Under sinusoidal inlet-pressure forcing conditions, the transverse jet does not preserve a fixed combustor structure. Instead, the wave system remains dynamically stabilized through bounded, phase-dependent switching between the previously identified combustion modes, with the triple-point trajectory remaining closed or bounded. Sufficiently large $J$ reverses this behavior: under the same forcing, the wave system develops drift-dominated motion and loses dynamic stabilization. The sampled stability map thus links the loss of dynamic stabilization to the combined effects of jet momentum and pressure-fluctuation amplitude, providing a quantitative design constraint that is absent from a purely steady-state analysis.

Three extensions merit investigation. The present simulations are two-dimensional, and three-dimensional computations are needed to resolve the spanwise instabilities and jet-in-crossflow dynamics that a planar model cannot capture. The dynamic-stabilization map should also be extended beyond single-frequency sinusoidal forcing to broadband, stochastic, and otherwise non-uniform inlet disturbances representative of realistic flight conditions. A complementary direction is to examine how the jet angle, injector number, and injector arrangement reshape the wave structure, thereby guiding transverse-injection strategies that broaden the stabilized operating window for liquid-fueled ODWE combustors.

## CrediT authorship contribution statement

**Wenhao Wang**: Writing – original draft, Visualization, Validation, Methodology, Investigation, Data curation. **Zongmin Hu**: Writing – review & editing, Supervision, Resources, Funding acquisition. **Peng Zhang**: Writing – review & editing, Supervision, Resources, Project administration, Funding acquisition, Conceptualization.

**Acknowledgments**

This work was supported by the National Natural Science Foundation of China (Grant No. 52176134 and 12172365). The work at the City University of Hong Kong was additionally supported by grants from the Research Grants Council of the Hong Kong Special Administrative Region, China (Project No. CityU 15222421 and CityU 15218820).

**Data availability:** Data will be made available on request.

**Declarations of interest:** none.